\documentclass[twocolumn]{aastex63}
\usepackage{epstopdf}
\usepackage{comment}
\usepackage{amsmath}
\shorttitle{A Superflare on RS CVn-type Star Triggered by \it SVOM}
\shortauthors{Wang et al.}

\begin{document}

\title{Multi-wavelength Study of A Superflare on RS CVn-type Star HD~22468 Triggered at Hard X-ray by \it SVOM}

\correspondingauthor{J. Wang, W. J. Xie}
\email{wj@nao.cas.cn, xiewj@nao.cas.cn}

\author{J. Wang} 
\affiliation{National Astronomical Observatories, Chinese Academy of Sciences, Beijing 100101,
People's Republic of China}

\author{W. J. Xie}
\affiliation{National Astronomical Observatories, Chinese Academy of Sciences, Beijing 100101,
People's Republic of China}

\author{F. Cangemi}
\affiliation{Universite Paris Cite, CNRS, Astroparticule et Cosmologie, F-75013 Paris, France}

\author{A. Coleiro}
\affiliation{Universite Paris Cite, CNRS, Astroparticule et Cosmologie, F-75013 Paris, France}

\author{H. L. Li}
\affiliation{National Astronomical Observatories, Chinese Academy of Sciences, Beijing 100101,
People's Republic of China}

\author{Y. Xu}
\affiliation{National Astronomical Observatories, Chinese Academy of Sciences, Beijing 100101,
People's Republic of China}

\author{X. H. Han}
\affiliation{National Astronomical Observatories, Chinese Academy of Sciences, Beijing 100101,
People's Republic of China}

\author{H. Yang}
\affiliation{Institute de Recherche en Astrophysique \& Planetologie, 9 avenue du colonel Roche, 31028 Toulouse Cedex 04, France}

\author{L. P. Xin}
\affiliation{National Astronomical Observatories, Chinese Academy of Sciences, Beijing 100101,
People's Republic of China}

\author{X. Mao}
\affiliation{National Astronomical Observatories, Chinese Academy of Sciences, Beijing 100101,
People's Republic of China}
\affiliation{School of Astronomy and Space Science, University of Chinese Academy of Sciences, Beijing, People's Republic of China}

\author{J. Zheng}
\affiliation{National Astronomical Observatories, Chinese Academy of Sciences, Beijing 100101,
People's Republic of China}

\author{J. J. Jin}
\affiliation{National Astronomical Observatories, Chinese Academy of Sciences, Beijing 100101,
People's Republic of China}

\author{G. W. Li}
\affiliation{National Astronomical Observatories, Chinese Academy of Sciences, Beijing 100101,
People's Republic of China}

\author{J. Rodriguez}
\affiliation{Universite Paris-Saclay, Universite Paris Cite, CEA, CNRS, AIM, 91191 Gif-sur-Yvette, France}

\author{L. Tao}
\affiliation{Institute of High Energy Physics, Chinese Academy of Sciences, 100049 Beijing}

\author{B. Cordier}
\affiliation{Universitee Paris-Saclay, Universitee Paris Cite, CEA, CNRS, AIM, 91191 Gif-sur-Yvette, France}

\author{J. Y. Wei}
\affiliation{National Astronomical Observatories, Chinese Academy of Sciences, Beijing 100101,
People's Republic of China}
\affiliation{School of Astronomy and Space Science, University of Chinese Academy of Sciences, Beijing, People's Republic of China}

\collaboration{22}{      }

\author{P. Bacon}
\affiliation{Universite Paris Cite, CNRS, Astroparticule et Cosmologie, F-75013 Paris, France}

\author{N. Bellemont}
\affiliation{Universite Paris Cite, CNRS, Astroparticule et Cosmologie, F-75013 Paris, France}

\author{L. Bouchet}
\affiliation{Institute de Recherche en Astrophysique \& Planetologie, 9 avenue du colonel Roche, 31028 Toulouse Cedex 04, France}

\author{H. B. Cai}
\affiliation{National Astronomical Observatories, Chinese Academy of Sciences, Beijing 100101,
People's Republic of China}

\author{C. Cavet}
\affiliation{Universite Paris Cite, CNRS, Astroparticule et Cosmologie, F-75013 Paris, France}

\author{Z. G. Dai}
\affiliation{Department of Astronomy, University of Science and Technology of China, Hefei 230026, China}

\author{O. Godet}
\affiliation{Institute de Recherche en Astrophysique \& Planetologie, 9 avenue du colonel Roche, 31028 Toulouse Cedex 04, France}

\author{A. Goldwurm}
\affiliation{Universite Paris Cite, CNRS, CEA, Astroparticule et Cosmologie, F-75013 Paris, France}
\affiliation{CEA Paris-Saclay, Irfu / Departement d'Astrophysique, F-91191 Gif-sur-Yvette, France}

\author{S. GUILLOT}
\affiliation{Institute de Recherche en Astrophysique \& Planetologie, 9 avenue du colonel Roche, 31028 Toulouse Cedex 04, France}

\author{L. Huang}
\affiliation{National Astronomical Observatories, Chinese Academy of Sciences, Beijing 100101, People's Republic of China}

\author{M. H. Huang}
\affiliation{National Astronomical Observatories, Chinese Academy of Sciences, Beijing 100101, People's Republic of China}

\author{N. Jiang}
\affiliation{Department of Astronomy, University of Science and Technology of China, Hefei 230026, China}

\author{E. W. Liang}
\affiliation{Guangxi Key Laboratory for Relativistic Astrophysics, School of Physical Science and Technology, Guangxi University, Nanning 530004, People's Republic of China}

\author{X. M. Lu}
\affiliation{National Astronomical Observatories, Chinese Academy of Sciences, Beijing 100101, People's Republic of China}

\author{S. Schanne}
\affiliation{Universite Paris-Saclay, Universite Paris Cite, CEA, CNRS, AIM, 91191 Gif-sur-Yvette, France}

\author{S. Le Stum}
\affiliation{Universite Paris Cite, CNRS, Astroparticule et Cosmologie, F-75013 Paris, France}

\author{Y. L. Qiu}
\affiliation{National Astronomical Observatories, Chinese Academy of Sciences, Beijing 100101, People's Republic of China}

\author{X. G. Wang}
\affiliation{Guangxi Key Laboratory for Relativistic Astrophysics, School of Physical Science and Technology, Guangxi University, Nanning 530004, People's Republic of China}

\author{X. Y. Wang}
\affiliation{School of Astronomy and Space Science, Nanjing University,  Nanjing 210023, Jiangsu, People's Republic of China}

\author{C. Wu}
\affiliation{National Astronomical Observatories, Chinese Academy of Sciences, Beijing 100101, People's Republic of China}

\author{L. Zhang}
\affiliation{Institute of High Energy Physics, Chinese Academy of Sciences, 100049 Beijing}

\author{S. N. Zhang}
\affiliation{Institute of High Energy Physics, Chinese Academy of Sciences, 100049 Beijing}

\author{S. J. Zheng}
\affiliation{Institute of High Energy Physics, Chinese Academy of Sciences, 100049 Beijing}






\begin{abstract}

Detection of stellar flares at hard X-ray is still rare at the current stage. 
A transient was recently detected by the hard X-ray camera, ECLAIRs onboard the 
SVOM mission at 11:39:01.2UT on 2025, January 09. Simultaneous monitor in the optical band on the ground by SVOM/GWAC and follow-up spectroscopy
enable us to confirm that the transient is caused by a superflare on HD~22468, a RS CVn-type star. 
The bolometric energy released in the flare is estimated to be 
$\sim7.2\times10^{37}-1.7\times10^{38}\ \mathrm{erg}$. 
The hard X-ray spectra of the event at the peak can be reproduced by the ``apec'' model of 
a hot plasma with a temperature of $106^{+27}_{-22}$~MK. In the optical range, 
the H$\alpha$ emission-line profile obtained at $\sim1.7$ hrs after the trigger shows a
bulk blueshift of $-96\pm20\ \mathrm{km\ s^{-1}}$, which can be explained by either a  
chromospheric evaporation or a prominence eruption. 
The ejected mass is estimated to be $3.9\times10^{20}$ g for the evaporating plasma, and to be
$3.2\times10^{21}\ \mathrm{g}<M_{\mathrm{p}}<8.8\times10^{21}\ \mathrm{g}$
for the erupted prominence.
\end{abstract}



\section{Introduction} \label{sec:intro}

Stellar flares, which release massive energy of $10^{33-39}\ \mathrm{erg}$ in a very short 
period of time, have been detected in multiple wavelengths, from radio to X-ray,
in main-sequence stars with a type from G to M, pre-main-sequence stars and RS CVn systems
(e.g., Pettersen 1989; Schmitt 1994; Osten et al. 2004, 2005; Huenemoerder et al.
2010; Maehara et al. 2012; Kowalski et al. 2013; Balona 2015;
Davenport et al. 2016; Notsu et al. 2016; Van Doorsselaere
et al. 2017; Chang et al. 2018; Paudel et al. 2018; Schmidt et al.
2019; Xin et al. 2021, 2024; Li et al. 2023b, Li et al. 2023a, 2024; Bai et al. 2023;
Wang et al. 2024; Mao et al. 2025). 
The RS CVn systems are close binary systems, which are usually composed of a (sub)giant and 
a dwarf or a subgiant.
There is accumulating evidence supporting that the
activity of host star is likely of decisive impact on  
the habitability of an exoplanet (e.g., Tian et al. 2011; Airapetian et al. 2016, 2017;
Cherenkov et al. 2017; Garcia-Sage et al. 2017;
Chen et al. 2021).

By analogy with the Sun, it is commonly accepted that the flares are generated by
magnetic reconnection (e.g., Noyes et al. 1984; Wright et al. 2011; Shulyak et al. 2017), 
although the strong magnetic fields might have diverse origins. The $\alpha\Omega$ dynamo is 
adopted to generate strong magnetic fields in G-type stars (e.g., Shibata \& Yokoyama 2002;
Getman et al. 2023), and the $\alpha^2$ dynamo in cool stars because of a lack of the boundary between the radiative and
convective zones (e.g., Hotta et al. 2022; Bhatia et al. 2023). 
Being different from the dynamos, the strong magnetic fields result from the tidal force between the binary stars in the RS CVn systems. 
Simon et al. (1980) proposed a scenario of 
a magnetic reconnection coupling the flux tubes of the two stars for the flares on the RS CVn systems.

Many details and open issues still need to be revealed and answered for the stellar flares.
On the one hand, unlike in the soft X-ray ($\mathrm{<10 keV}$) band,  
the hard X-ray ($\mathrm{\gg10 keV}$) emission is still rarely detected in the flares of 
remote stars, partially because of the limited sensitivity of the hard X-ray survey facilities. 
Two superflares on II Peg and EV Lac detected by \it Swift\rm/BAT have been reported and 
studied in Osten et al. (2007) and Osten et al. (2010), respectively. 
Tsuboi et al. (2016) reported 23 superflares on 13 active stars detected in the 2-30~keV
range by the first two years survey of Monitor of All-sky X-ray Image (MAXI). By analyzing the archived data of the Nuclear spectroscopic
Telescope Array (NuSTAR), NuSTAR~J230059+5857.4 is identified as a stellar
superflare resulting from a plasma with a temperature of 95~MK (Hakamata et al. 2025).

On the other hand, the insufficient spatial resolution of contemporary instruments
results in a difficulty in detecting complicated dynamics, e.g., 
coronal mass ejection (CME) and chromospheric evaporation, caused by magnetic reconnection 
in distant stars (see Leitzinger \& Odert (2022) and references therein for a recent
review). A batch of candidates of CME or chromospheric evaporation (or prominence eruption) 
has been recently identified based on either asymmetry or bulk velocity shift of the 
H$\alpha$ and \ion{O}{8}$\lambda18.97$\AA\ emission lines (e.g., Moschou et al. 2019; Argiroffi et al. 2019; 
Cao \& Gu 2024; Wang et al. 2021, 2022, 2024; Wu et al. 2022; Chen et al. 2022; Namekata et al. 2021, 2024).
A filament eruption has recently been identified in the RS CVn-type star UX AI by the blueshifted H$\alpha$
absorption with a bulk velocity of $\sim140\ \mathrm{km\ s^{-1}}$ (Cao \& Gu 2025).

In this paper, we report the discovery and a follow-up observation in time-resolved spectroscopy for the hard 
X-ray transient SVOM~J00365+0033 triggered by SVOM/ECLAIRs, which allows us to identify that the transient is caused by a flare from a RS CVn-type star. SVOM, launched on 2024 June 22, is a Chinese-French
space mission dedicated to the detection and study of gamma-ray bursts and high-energy transients.
We refer the reader to Atteia et al. (2022) and the white paper
given by Wei et al. (2016) for details.
The follow-up spectroscopy reveals a low-speed outflow possibly related to 
a chromospheric evaporation or a prominence eruption associated with the flare.

The paper is organized as follows. Section 2 presents the hard X-ray transient triggered by the SVOM/ECLAIRs.
A simultaneous monitor by SVOM/GWAC on the ground is described in Section 3.  
Our follow-up observations in spectroscopy, along with data reduction, are given in Section 4. 
Section 5 shows the analysis and results. The implications are presented in Section 6.


\section{Hard X-ray Transient Detected by SVOM/ECLAIRs} \label{sec:style}
 
The hard X-ray transient SVOM~J00365+0033 (ID$=$sb25010902) triggered the
soft $\gamma$-ray Camera ECLAIRs on board SVOM at 2025-01-09UT11:39:01.2 (MJD~$=$~60684.98543, hereafter $t_0$).

ECLAIRs is the wide-field (FoV$\sim$2 sr) coded-mask ($\sim40\%$ transparency
below 80 keV) trigger camera of SVOM 
(e.g., Schanne et al. 2013; Le Provost et al. 2013; Godet et al. 2014; Lacombe et al. 2014; Xie et al. 2024; Llamas Lanza et al. 2024). It works in the energy range of 4-120 keV, and has 
an active area $\approx1000\ \mathrm{cm^{2}}$ in the detection plane. 
The localization accuracy is 13\arcmin\ at a 90\% confidence level for on-axis 
weak sources with a signal-to-noise ratio (S/N) $\mathrm{\sim10}$, and  
better than 2\arcmin\ for bright sources with $\mathrm{S/N\sim100}$ (Goldwurm et al. 2026, 
in preparation). 

\begin{figure}
    \centering
    \plotone{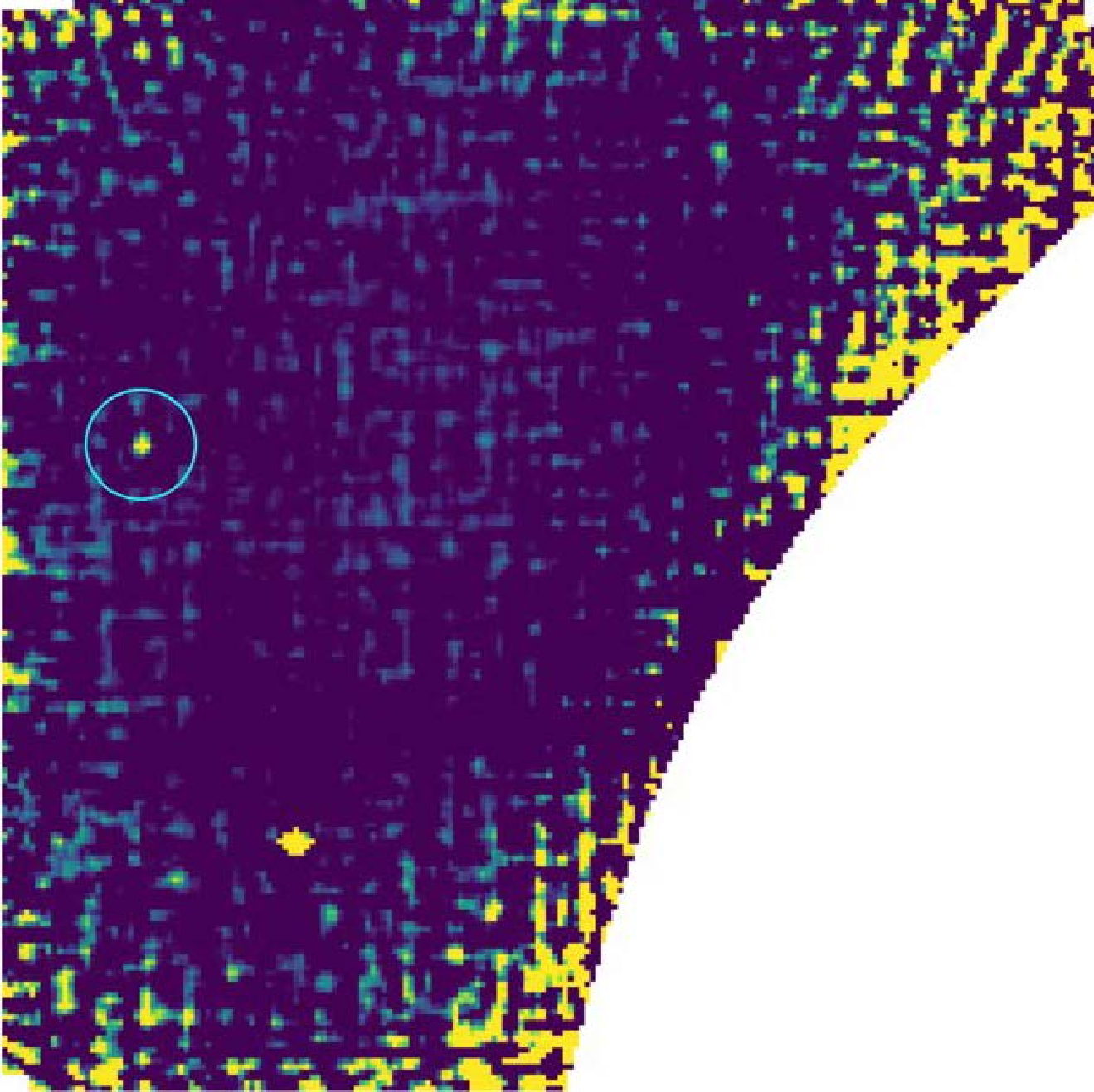}
    \caption{The discovery image of SVOM~J00365+0033 (ID$=$sb25010902) taken by SVOM/ECLAIRs. The source is marked by the blue circle. The bright source at the
    bottom right of SVOM~J00365+0033 is the Crab. The blank at the right-bottom corner is due to the obscuration of the Earth.
    \label{fig:discovery}}
\end{figure}

Figure \ref{fig:discovery} shows the discovery sky image of the transient taken by ECLAIRs.
The transient with an S/N of 9.47 was located at R.A.=
$\mathrm{03^h36^m39^s}$
and Dec=$\mathrm{00\degr33\arcmin25\arcsec}$ with an uncertainty of 8\arcmin.4 (radius, 90\%\ confidence level 
including both statistical and systematic errors) based on Eq.(10) in Goldwurm \& Gros (2022).

At a distance of 4\arcmin\ away from the center determined by ECLAIRs,
there exists a bright X-ray source listed in the ROSAT All-sky Bright Source Catalog
(Voges et al. 1999),
enhanced 3XMM catalog (Rosen et al. 2016), and 
Chandra source catalog (Evans et al. 2010).
The X-ray source (1RXS~J033647.2+003518 = 3XMM~J033647.2+003515) has a flux of $7.9\times10^{-11}\ \mathrm{erg\ s^{-1}\ cm^{-2}}$ in the 0.2--12 keV band (Rosen et al. 2016), and shows X-ray flares detected by 
XMM-Newton serendipitous observations as reported in Pye et al. (2015). 
This X-ray source is associated with a RS CVn-type star HD~22468 ($=$HR~1099) that is 
well known for its plentiful flares from radio to X-ray 
(e.g., Foing et al. 1994; Osten et al. 2004; Pandey \& Singh 2012; Didel et al. 2025).
Combining these facts and the evolution in X-ray, optical continuum and H$\alpha$ 
emission line (see below) enable us to identify the event SVOM~J00365+0033 
as resulting from a 
stellar flare occurring on HD~22468. The basic properties of HD~22468 are listed in Table \ref{tab:properties}. 

\begin{table}
        \centering
        \caption{Properties of HD~22468}
        \footnotesize
        \label{tab:example_table}
        \begin{tabular}{cc} 
        \hline
        \hline
        Property & Value \\
          (1) & (2)  \\     
        \hline 
       R.A. &  $\mathrm{03^h:36^m:47^s.3}$\\
       Dec  &  +00\degr:35\arcmin:16\arcsec\\
       Sp. T & K1IV+G5V\\
       Distance (pc) & $29.43\pm0.03$\\
       $M_G$ (mag) & 2.9\\
       $P_{\mathrm{rot}}$ (day) & 2.84 \\
       $V\sin i$ ($\mathrm{km\ s^{-1}}$) & 34.5 \\
       \hline
       \multicolumn{2}{c}{Primary}\\
       \hline
       G-band (mag) & $5.62$\\
       GBp-GRp (mag) & $1.22$  \\
       $M_{\star}$ ($M_\odot$)  &  1.5 \\
       $R_\star$ ($R_\odot$) & 4.0 \\
       $T_{\mathrm{eff}}$ (K) & $4712\pm38$\\
       $\log(g/\mathrm{cm\ s^{-2}})$ & $3.4\pm0.1$\\
       $\mathrm{[Fe/H]}$ &  $-0.16\pm0.05$\\
       \hline
       \multicolumn{2}{c}{Secondary}\\
       \hline
       G-band (mag) & $8.52$\\
       GBp-GRp (mag) & $1.19$  \\
       $M_{\star}$ ($M_\odot$)  &  0.6 \\
       $T_{\mathrm{eff}}$ (K) & $4631\pm50$\\
       $\log(g/\mathrm{cm\ s^{-2}})$ & $4.5\pm0.1$\\
        \hline
        \end{tabular}
        \tablecomments{References: Gaia Collaboration et al. (2022); Kervella et al. (2022); Soubiran et al. (2022); Seli et al. (2022), Luck et al. (2017)}
        \label{tab:properties}
\end{table}

Even though the high S/N ratio of 9.47 of the event was reached, 
a follow-up  snapshot by both the narrow-field instruments, MXT and VT, onboard SVOM 
was not performed because an automatic slew of the platform has been rejected due to an
ongoing Target-of-Opportunity observation with a higher priority. The star is actually too bright to be observed by 
VT. A star with a brightness of 8th magnitude is in fact saturated in the VT image for the shortest exposure
of 1 second. 


\section{Simultaneous Monitor by SVOM/GWAC and Data Reduction}

During the night of 2025, January 09, the sky area covered by the ECLAIRs was simultaneously monitored by the SVOM/GWAC 
(Ground-based Wide Angle Cameras) deployed in Xinglong Observatory, Chinese Academy of Sciences, since 
the observation plan of SVOM/GWAC is designed to monitor the sky region covered by ECLAIRs as much as possible (see Han et al. 2021, Li et al. 2024 for a description of GWAC in
details). 

Up to the beginning of 2025, SVOM/GWAC was in total composed of 10 mounts. Each mount has  
four Joint Field-of-View (JFoV) cameras and one Full Field-of-View (FFoV) cameras. 
The diameters of the JFoV and FFoV cameras are 18 cm and 3.5 cm, respectively.
The FFoV is used to guide the pointing of the mount, and to monitor the bright sources 
that are saturated in the JFoV images. 
A 4K $\times$ 4K Dhyana 4040BSI CMOS detector is equipped on each JFoV camera, and the ZWO ASI2600MCAir 
CMOS\footnote{https://www.zwoastro.com/product/asi2600/} detector, 
with a size of $6248\times4176$ pixels, on each FFoV camera. 
The field-of-view (FoV) is $9\degr.8\times9\degr.8$ for each JFoV camera, and $64\degr\times42\degr$ for each FFoV camera. The total FoV of the 40 JFoV cameras is $\sim3600\ \mathrm{deg^2}$.  
The exposure time of each camera is fixed to be 3~s, and the readout time is 1~s, which result in a cadence of 4~s.

The sky field around the event of SVOM~J00365+0033 was monitored by the GWAC mount \#2
from MJD~60684.92381 to 60685.095429, i.e., $\approx4.11$~hr duration.

The host star HD~22468 was in fact saturated in the JFoV images, but not in the images taken 
by the FFoV camera. 
The FFoV images were reduced by the standard procedure, including
bias, dark, and flat-field corrections, by using the IRAF\footnote{IRAF is distributed by the National Optical Astronomical Observatories,
which are operated by the Association of Universities for Research in
Astronomy, Inc., under cooperative agreement with the National Science
Foundation.} package. An aperture photometry was then performed by adopting an aperture size of 3 pixels. Finally, the USNO B1.0 catalog (Monet et al. 2003) was used to carry out 
absolute photometric calibration. The limiting magnitude of FFoV is determined to be 11.8 mag during the monitor.

\section{Spectroscopic Follow-up Observations and Data Reductions}

Spectroscopy of HD~22468 has been taken with the NAOC 2.16 m telescope 
(Fan et al. 2016) in three observational runs after the ECLAIRs trigger.
The log of the spectroscopic observations is tabulated in Table \ref{tab:spec_log}, where the Column (1) lists the 
start time of each observational run. Our first spectrum was obtained
$\approx1.7$~hr after the trigger. 

\begin{table*}
\renewcommand{\thetable}{\arabic{table}}
\centering
\caption{Observational Log and Results of Optical Spectroscopic Analysis.}
\label{tab:spec_log}
\begin{tabular}{cccccc}
\tablewidth{0pt}
\hline
\hline
UT & Exp time & $f_\mathrm{c}$ &  $\mathrm{EW(FeI\lambda6495)}$ & $f_\mathrm{H\alpha}$  & 
$\Delta\upsilon(\mathrm{H\alpha})$ \\
 & second  & $\mathrm{erg\ s^{-1}\ cm^{-2}\ \AA^{-1}}$ & $\mathrm{\AA}$ & $\mathrm{erg\ s^{-1}\ cm^{-2}}$ &   $\mathrm{km\ s^{-1}}$ \\
(1)  &   (2) & (3) & (4) & (5) & (6) \\
\hline
2025-01-09T13:22:55 &  $10\times30$  & $1.6\times10^{-11}$  & $1.10\pm0.09$ & $(2.4\pm0.1)\times10^{-11}$   & $-96\pm20$ \\
2025-01-10T11:11:45 &  $10\times30$  & $1.4\times10^{-11}$  & $0.97\pm0.07$ & $(1.8\pm0.1)\times10^{-11}$   & $-50\pm16$ \\
2025-02-05T10:55:35 &  $10\times30$  & $1.3\times10^{-11}$  & $0.99\pm0.08$ & $(1.1\pm0.1)\times10^{-11}$ &  $-14\pm17$\\
\hline
\hline
\end{tabular}
\tablecomments{Column (1): the start time of spectroscopy in UT. Column (2): the total exposure time in unit of second. 
Column (3): the continuum flux density measured in the wavelength range between 6515\AA\ and 6545\AA.
Column (4): the equivalent width of \ion{Fe}{1}$\lambda6496$ absorption feature.
Column (5): the flux of H$\alpha$ emission line. 
Column (6): the bulk velocity shift of H$\alpha$ line measured with respect to the stellar photosphere.}
\end{table*}

All spectra were taken
with the Beijing Faint Object Spectrograph and Camera
that is equipped with a back-illuminated E2V55-30 AIMO CCD. 
A total of 10 frames were obtained in each observational run, in which
the exposure time of each frame is 30 seconds. 
The G8 grism with a wavelength coverage of 5800 to 8200\AA\ was used in the
observations, which allows us to study the H$\alpha$ emission-line profile with 
adequate spectral resolution. With a slit width of 1\arcsec.8 oriented in the 
south–north direction, the resolution is measured to be 3.5\AA\ 
according to the sky emission lines, 
which corresponds to $R=\lambda/\Delta\lambda=1880$ and a velocity of $160\ \mathrm{km\ s^{-1}}$ for the H$\alpha$ emission line. The wavelength calibration was carried out with iron–argon comparison
lamps. Flux calibration was carried out by the
observations of the standard stars from the Kitt Peak National Observatory (Massey et al. 1988).

The one-dimensional (1D) spectrum was
extracted from each of the raw images by using the IRAF package
and standard procedures, including bias subtraction and flat-field correction.
In the spectral extraction, the apertures of both source and sky
emission were fixed for both object and corresponding standard.
The extracted 1D spectra were
then calibrated in wavelength and in flux by the corresponding
comparison lamps and standard stars, respectively. The zero-point of the
wavelength calibration was corrected for each spectrum by using 
the sky [\ion{O}{1}]$\lambda$6300 emission line as a reference. 
The accuracy of wavelength calibration is therefore determined to be $\sim0.1$\AA, 
which corresponds to a velocity of $\sim5\ \mathrm{km\ s^{-1}}$ for the H$\alpha$ line.


\section{Analysis and Results}

\subsection{Hard X-ray Spectra and Light Curves}

After taking into account the obscuration by the Earth, 
the hard X-ray spectra of SVOM~J00365+0033
are extracted in four epochs, i.e., $[t_0-4000,\ t_0-3000]$~s, $[t_0,\ t_0+940]$~s, $[t_0+1444,\ t_0+3500]$~s and $[t_0+7273,\ t_0+9300]$~s, 
by the dedicated pipelines of ECLAIRs\footnote{https://fsc.svom.org/documentation/ecpi/index.html} and the corresponding 
calibration files, where $t_0$ refers to the trigger time.  

\begin{figure*}
   \begin{tabular}{ll}
   \begin{minipage}{0.48\linewidth}
      \centering
      \plotone{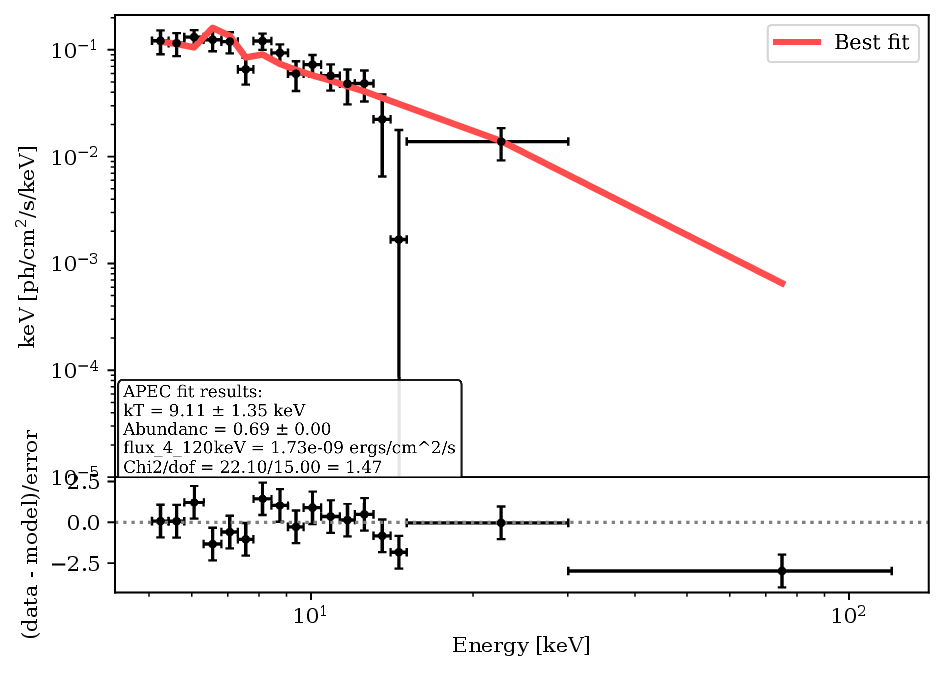} 
   \end{minipage} & 
   \begin{minipage}{0.48\linewidth}
      \centering
      \plotone{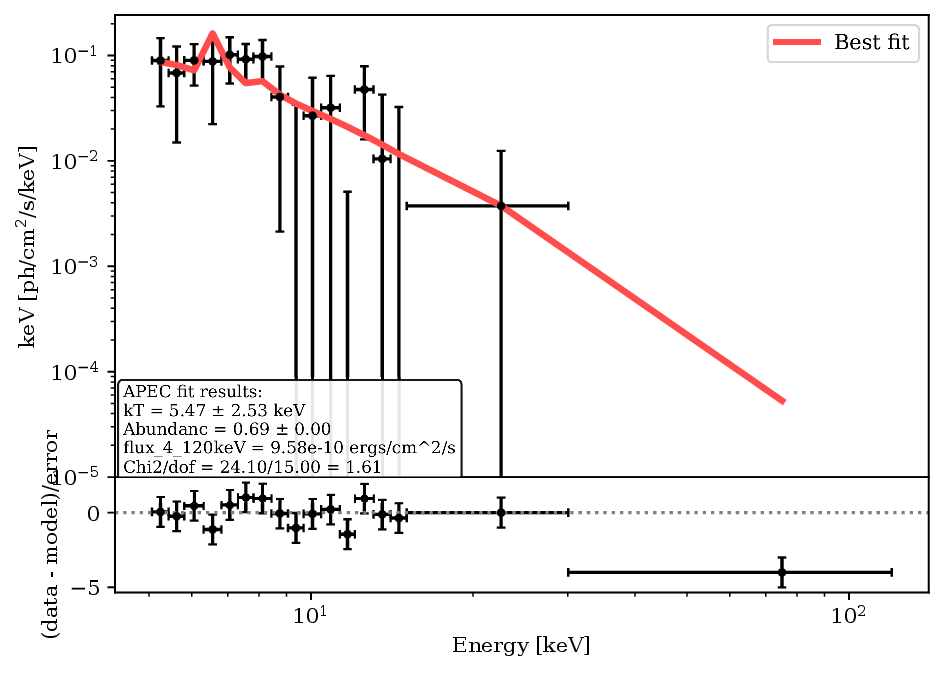}
   \end{minipage}  \\
   \begin{minipage}{0.48\linewidth}
      \centering
      \plotone{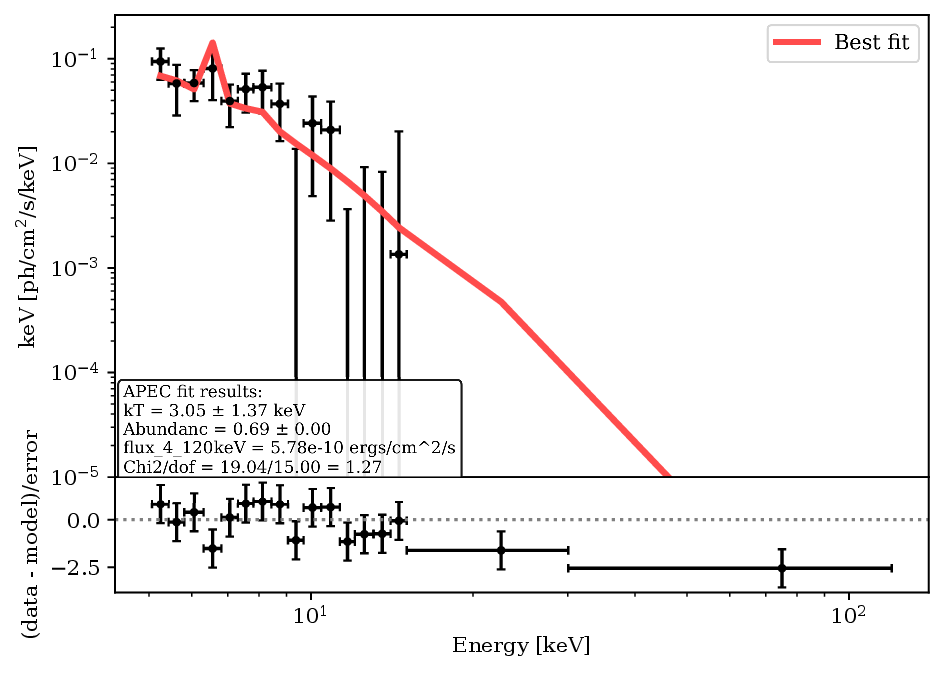}
   \end{minipage} \\
   \end{tabular}

   \caption{
   \it Upper-left panel: \rm Hard X-ray spectrum of SVOM~J00365+0033 observed by SVOM/ECLAIRs in the epoch of
   $[t_0,\ t_0+940]$s, where $t_0$ is the trigger time. The best-fit \tt apec \rm model is overplotted by the red line. 
   A fixed abundance of $0.69Z_\odot$ is adopted in the modeling.
   The subpanel underneath the spectrum shows the residuals, in units of 
   $\mathrm{counts\ s^{-1}\ keV^{-1}}$, of the observed data from the best-fit model.
   \it Upper-right and lower-left panels: \rm the same as the left one, but for the epochs of
   $[t_0+1444,\ t_0+3500]$s and $[t_0+7273,\ t_0+9300]$s, respectively. 
}
\label{fig: xspec}
\end{figure*}

To investigate the evolution of its hard X-ray emission,  
we model the four time-integrated hard X-ray spectra by using the Xspec software
(version 12.14.1, Arnaud 1996). 
The \tt apec \rm  model (e.g., Smith et al. 2001) 
are adopted to reproduce the observed spectra. 
The abundance is fixed to be 0.69$Z_\odot$ in the modeling\footnote{The abundance is inferred from the 
[Fe/H] value by assuming the solar composition.}. 
The best fits are illustrated in Figure \ref{fig: xspec} for the three epochs after the  trigger, and the corresponding results are 
tabulated in Table \ref{tab:xproperties}.
All the errors reported in the table are obtained by using the \tt error \rm command in the 
Xspec package, and are at the 90\% significance level.

In the second epoch, the spectrum can be reproduced by
a hot plasma with a temperature of $106^{+27}_{-22}$~MK when
the \tt apec \rm model is adopted (see the upper-left panel in Figure \ref{fig: xspec}). This temperature is actually comparable to the measurement in the stellar flare event NuSTAR~J230059+5857.4 (Hakamata et al. 2025).

\begin{table}
\centering
        \caption{X-Ray Spectral Fit Parameters of SVOM~J00365+0033}
        \label{tab:example_table}
        \footnotesize
        \begin{tabular}{ccccc} 
         \hline
         \hline
         $t-t_0$ & $kT$ & Norm & $F_{\mathrm{4–120 keV}}$ & $\chi^2/\mathrm{d.o.f}$ \\
         second & keV &     & $\mathrm{erg\ cm^{-2}\ s^{-1}}$ & \\
           (1) & (2) & (3) & (4) & (5)\\     
         \hline 
         $(-4000, -3000)$ & $7.8^{+9,42}_{-4.92}$ & $0.66_{-0.29}^{+1.21}$ & $8.0_{+1.4}^{-4.4}\times10^{-10}$ & 1.08\\
         $(0, 940)$ & $9.2_{-1.9}^{+2.3}$ & $1.2_{-0.2}^{+0.2}$ & $1.7_{-0.2}^{+0.1}\times10^{-9}$ & 1.47\\ 
         $(1444, 3500)$ & $4.7^{+2.6}_{-1.9}$ & $1.3^{+1.4}_{-0.5}$ & $9.2_{-3.1}^{+0.6}\times10^{-10}$ & 1.61 \\
         $(7273, 9300)$ & $3.0^{+2.5}_{-1.6}$ & $1.6^{+6.8}_{-0.9}$ & $5.8_{-4.2}^{+0.2}\times10^{-10}$ & 1.27 \\
         \hline
         \end{tabular}
         \tablecomments{Column (1):  The time window in which the X-ray spectra are extracted and modeled, where $t_0$ is the trigger time of ECLAIRs.  Column (2-4): The 
         best-fit temperature, normalization and flux in $4-120$~keV when the \tt apec\rm\ model is adopted. The reported uncertainties correspond to a 90\% 
         significance level. Column (5): The final reduced $\chi^2$.}
         \label{tab:xproperties}
\end{table}

\begin{figure*}
    \centering
    \plotone{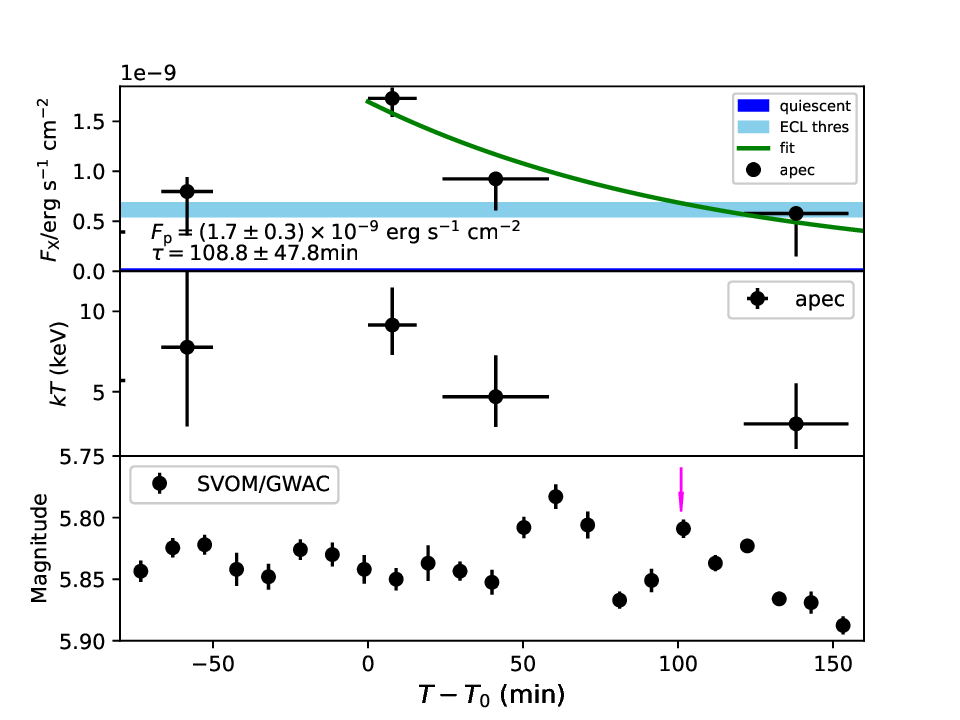}
    \caption{\it Top panel: \rm the hard X-ray light curve of SVOM~J00365+0033 detected by SVOM/ECLAIRs in $4-120$~keV band. The dark-blue shadow region at the bottom of the panel 
    marks the quiescent emission level estimated from the 3XMM catalog (see main text 
    for the details). The sky-blue region corresponds to the $3\sigma$ ECLAIRs detection thresholds for a 1~000 seconds exposure. $T_0$ is the ECLAIRs trigger time.  \it Middle panel: \rm evolution of the modeled plasma temperature.
    \it Bottom panel: \rm the white-light
    light curve of HD~22468 monitored by the SVOM/GWAC FFoV camera in 2025, January 09, 
    after binning the data by averaging the measurements within every 10 minutes. 
    The epoch of our first spectroscopic follow-up is marked by the downward magenta arrow. 
    \label{fig: evolution}}
\end{figure*}

After the modeling of the energy spectra, the evolution of the modeled X-ray flux in the 
$4-120$~keV band and the plasma temperature are displayed in the upper two panels of 
Figure \ref{fig: evolution}. In the top panel, the quiescent flux level in the $4-120$~keV band, i.e., 
$F_{\mathrm{X,q}}=5.6\times10^{-13}-1.7\times10^{-11}\ \mathrm{erg\ s^{-1}\ cm^{-2}}$, is converted 
from the flux listed in the 3XMM catalog by using the \tt pimms \rm task, in which 
the \tt apec \rm model is considered. The plasma 
temperature ranges from $kT=1$~keV to 3~keV in the conversion. 
The ECLAIRs detection thresholds\footnote{The threshold at a 3$\sigma$ significance level is estimated to have a count rate of 
$0.07\ \mathrm{count\ s^{-1}}$ for an 1~000 seconds exposure.} at a 3$\sigma$ significance level are calculated 
by adopting the same models, and marked in the plot by the light-blue region. 

As shown in the plot, the event was marginally detected before the trigger time. 
After the trigger time,
the light curves are typical of a stellar flare in which the hard X-ray emission 
quickly fades out along with a fast cooling of the plasma (e.g., Mao et al. 2025). 
We model the hard X-ray light curve by 

\begin{equation}
 F(t) = (F_{\mathrm{p}}-F_{\mathrm{q}})\times\exp{\bigg(-\frac{t-t_{\mathrm{p}}}{\tau}\bigg)}+F_{\mathrm{q}},\ t>t_{\mathrm{p}}
\end{equation}
where $F_{\mathrm{p}}$ and $t_{\mathrm{p}}$ are the peak flux and time, respectively. 
$F_{\rm q}$ is the corresponding quiescent flux.
By setting $t_{\mathrm{p}}=t_0$, the fitting 
yields $F_{\mathrm{p}} = (1.7\pm0.3)\times10^{-9}\ \mathrm{erg\ s^{-1}\ cm^{-2}}$ and $\tau = 108.8\pm 47.8$~min.
The best fit is overploted in Figure \ref{fig: evolution} by the solid line. 
Based on the fitting, the total energy released in the 
$4-120$~keV is roughly estimated to be 
$E_{\mathrm{4-120keV}} = (1.2\pm0.5)\times10^{36}\ \mathrm{erg}$ for SVOM~J00365+0033,
which corresponds to a value of $E_{\mathrm{X}}\sim1.7\times10^{36}\ 
\mathrm{erg}$ in the $0.2-12$~keV band by using the \tt pimms \rm task\footnote{The plasma temperature is fixed to be $kT=9.2$~keV.}. The bolometric energy released in the flare 
is then estimated to be $E_{\mathrm{bol}}\sim1.7\times10^{38}\ \mathrm{erg}$,
when a bolometric correction $E_{\mathrm{X}}/E_{\mathrm{bol}}=0.01$ is adopted (e.g., Emslie et al. 2012; Wang 2023).  
 
The emission measurement (EM) is 
calculated to be $\mathrm{EM}=\int n_{\mathrm{e}}^2 \mathrm{d}V=4\times10^{14}\pi d^2\times\mathrm{norm}=(1.3\pm0.2)\times10^{55}\ 
\mathrm{cm^{-3}}$, where $n_{\mathrm{e}}$ is the electron density, $V$ the emitting volume, and $d$ the distance to the star in unit of cm. The
norm is one of the parameters in the \tt apec \rm model.

%
%

\subsection{Light Curve in White-light}

After binning with a bin size of 10 minutes, the light curve derived from the GWAC images is presented in the bottom panel in Figure 3 for HD~22468.
A white-light (WL) flare peaking at $\sim1$ hour after the ECLAIRs trigger can be 
identified for 
a brightening of $\sim0.05$ mag at a significance level of $\sim3\sigma$, which supports the claim 
that SVOM~J00365+0033 originated from stellar activity on HD~22468. 

The flaring energy released in the $R-$band is estimated to be $E_R\sim1.2\times10^{37}$~erg 
by an integration of the light curve. This value yields a bolometric energy of
$E_{\mathrm{bol}}\sim7.2\times10^{37}$~erg when a bolometric correction 
of $E_{\mathrm{bol}}/E_R=6$ derived from a blackbody with a temperature of $10^4$~K is adopted. 
The effect of reddening caused by the Galaxy is ignored in the calculation due to the extremely small 
color excess $E(B-V)\approx0.02$~mag. 
By assuming the hydrogen  density around the Sun of $n_{\mathrm{H}}=10^6\ \mathrm{cm^{-3}}$ and 
the constant dust-to-gas ratio, this color excess is estimated from the calibration of 
$E(B-V)\approx0.53\times(d/\mathrm{kpc})$ (Bohlin et al. 1978), where $d$ is the distance of the object.

\subsection{Decay and Blueshift of H$\alpha$ Emission Line}

\begin{figure*}
    \centering
    \plotone{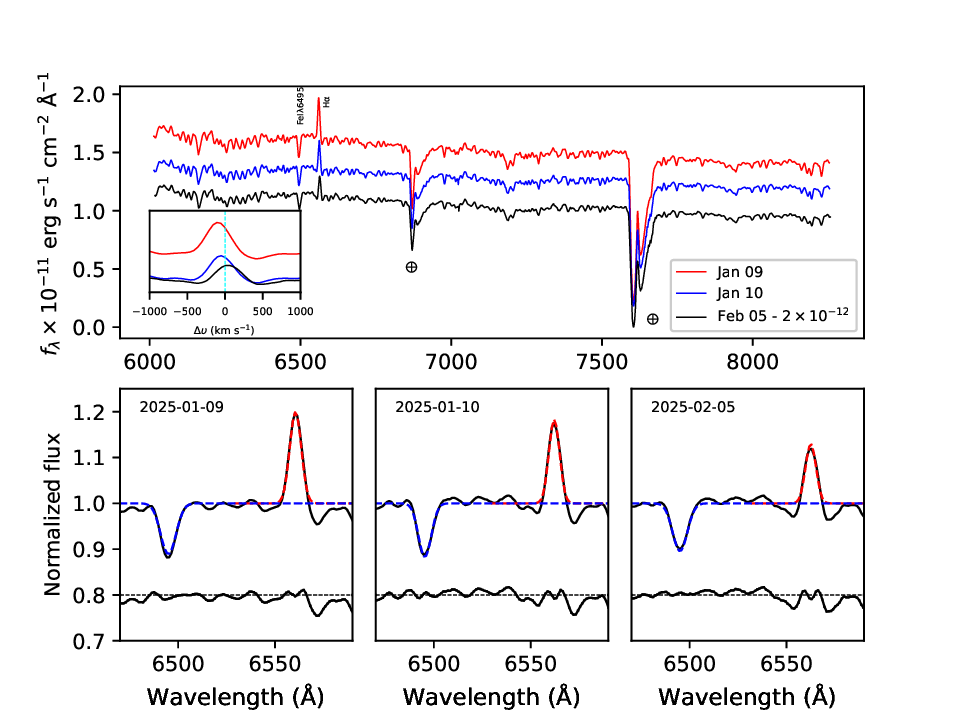}
    \caption{\it Upper panel: \rm a comparison of the spectra of HD~22468 taken in the three 
    different epochs. Note that the spectrum in the quiescent state (black line) obtained in 2025, 
    February 05
    is shifted vertically by an arbitrary amount for visibility. The insert panel compares the H$\alpha$ 
    line profiles obtained in the three different epochs. \it Bottom panels: \rm the modeling of the 
    \ion{Fe}{1}$\lambda$6496 absorption feature and the H$\alpha$ emission line in 
    the three epochs. Each line is reproduced
    by a Gaussian function. In each sub-panel, the observed and modeled line profiles are plotted by 
    the black and red (blue) solid lines, respectively. The curves underneath each line spectrum 
    present the residuals between the observed and modeled profiles.
    \label{fig:spec}}
\end{figure*}

After combining the individual spectra taken in each run, the three combined spectra taken at the three
different epochs are compared in the upper panel of Figure \ref{fig:spec}. In addition to the 
\ion{Fe}{1}$\lambda$6496 absorption feature arising from the photosphere, a decrease 
can be observed in both continuum and H$\alpha$ emission line from the chromosphere.

After normalizing the continuum by a spline function, each of the features of \ion{Fe}{1}$\lambda$6496 
and H$\alpha$ is modeled by a Gaussian function by the IRAF/SPECFIT task (Kriss 1994).  
The modeling is illustrated in the three lower panels in Figure 4. The results of the modeling are 
tabulated in Table \ref{tab:spec_log}, where all the uncertainties correspond to the 1$\sigma$ significance level resulted from
the modeling. 
$f_{\mathrm{c}}$ listed in Column (3) is the continuum flux level directly measured in the wavelength range from 6515\AA\ to 6545\AA\ by the \tt 
splot \rm task in the IRAF package.
The continuum slightly decreases from $1.6\times10^{-11}\ \mathrm{erg\ s^{-1}\ cm^{-2}\ \AA^{-1}}$ 
to the quiescent level of
$1.3\times10^{-11}\ \mathrm{erg\ s^{-1}\ cm^{-2}\ \AA^{-1}}$, which corresponds to a
differential magnitude of $\sim 0.2$ mag. In addition to the decay of the continuum, 
the H$\alpha$ line emission decreases by a factor of
two from the active to the quiescent states. The decay of both continuum and H$\alpha$ line emission reinforces our identification that 
SVOM~J00365+0033 is resulting from stellar activity on RS CVn-type star HD~22468.

Column (6) in Table~\ref{tab:spec_log} lists the bulk velocity shift of the H$\alpha$ emission 
$\Delta\upsilon_{\mathrm{H\alpha}}$. It is calculated as

\begin{equation}
 \Delta\upsilon_{\mathrm{H\alpha}}=c\times\frac{\Delta\lambda-\Delta\lambda^0}{\lambda^0_{\mathrm{H_\alpha}}}
\end{equation}
where $\Delta\lambda=\lambda_{\mathrm{H\alpha}}-\lambda_{\mathrm{FeI}}$ and 
$\Delta\lambda^0=\lambda_{\mathrm{H\alpha}}^0-\lambda_{\mathrm{FeI}}^0$.
$\lambda^0_{\mathrm{H\alpha}}$ ($\lambda_{\mathrm{H\alpha}}$) and $\lambda^0_{\mathrm{FeI}}$
($\lambda_{\mathrm{FeI}}$) are the (observed) wavelengths  
measured in laboratory of the H$\alpha$ and \ion{Fe}{1}$\lambda$6495 lines, respectively. \rm According to this definition, a positive $\Delta\upsilon_{\mathrm{H\alpha}}$ denotes a downflow, and a negative one an upflow.

We argue that the large H$\alpha$ off-set velocity 
of $\sim-100\ \mathrm{km\ s^{-1}}$ revealed in the
spectrum taken at $\sim1.7$ hrs after the trigger of ECLAIRs
is most likely resulted from an upward plasma motion related with the flare. 
On the one hand, since the absorption \ion{Fe}{1} feature certainly comes from the 
underlying photosphere, the velocity $\Delta\upsilon_{\mathrm{H\alpha}}$ is therefore 
unaffected by the orbiting of the binary, and reflects a plasma
motion with respect to photosphere. On the other hand, the contribution from chromospheric
active region associated with stellar rotation is too small to explain the measured large 
H$\alpha$ off-set velocity. In fact, the rotational velocities have been measured to be 
$\sim41\ \mathrm{km\ s^{-1}}$ and $\sim8\ \mathrm{km\ s^{-1}}$ for the cool and hot stars, 
respectively (De Medeiros et al. 2004; Glebocki \& Gnacinski 2005; Herrero et al. 2012; Luck 2017).


\section{Discussion}

A multi-wavelength study is performed on the transient event SVOM~J00365+0033 triggered by 
SVOM/ECLAIRs in 2025, January 09, which enables us to identify the event is originated from 
a flare from RS CVn-type star HD~22468. The star was revealed to have both a WL flare and enhanced H$\alpha$ emission seen in optical band at $1-2$ hours after the trigger.  
The bolometric energy released in the flare is estimated to be 
$\sim7.2\times10^{37}\ \mathrm{erg}$ and $\sim1.7\times10^{38}\ \mathrm{erg}$ from the WL and 
X-ray light curves, respectively. \rm In addition, an upflow with 
a velocity of $-96\pm20\ \mathrm{km\ s^{-1}}$ with respect to the photosphere is revealed from 
the H$\alpha$ emission line.

\subsection{Discrepancy in the Bolometric Energy}

Although an estimation of bolometric energy strongly depends on the adopted bolometric correction, 
the above discrepancy between the bolometric energy obtained from the light curves in the two different bands is likely resulted from the different integration ways.
Not as a modeling by an exponential decay in the X-ray band,
an over simplified integration is adopted for the WL light curve, which result in an
underestimation of the total released energy as follows. 
Because of the small flaring amplitude in WL, 
the WL light curve is integrated by simply adding all 
the points exceeding the 1$\sigma$ fluctuation determined in the quiescent state before the 
flare. The relatively large fluctuation naturally leads to an neglecting of the late gradual 
decaying phase in the integration. The gradual decaying phase has non-negligible contribution to 
the total released energy since its long duration, even though at a low emission level.

\rm
\subsection{Flare Properties}

\begin{figure*}
    \centering
    \plotone{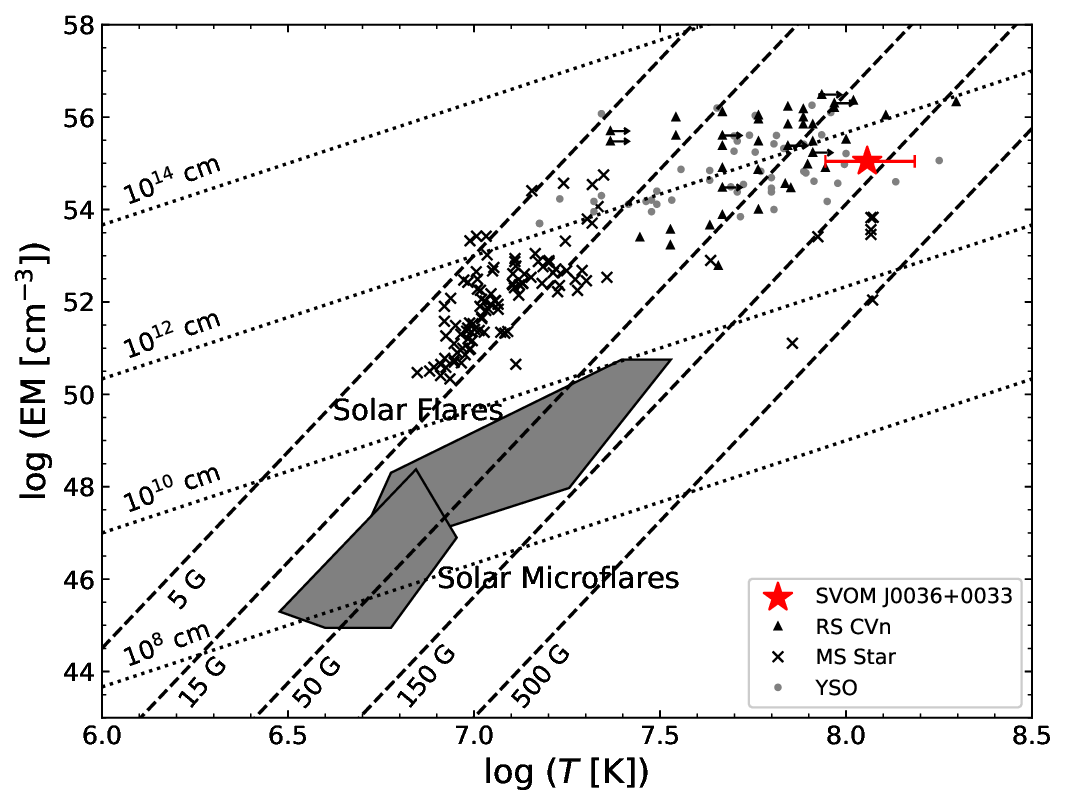}
    \caption{EM$-T$ diagram. The event SVOM~J00365+0033 is marked with the red star.
    The triangles show the RS CVn-type stars (Tsuru et al. 1989; Endl et al. 1997; Franciosini et al. 2001; Pandey \& Singh 2012; Tsuboi et al. 2016; Sasaki et al. 2021; Karmakar et al. 2023; Mao et al. 2025), and the crosses the main-sequence stars (Pye et al. 2015). The gray dots mark the positions
    of young stellar objects extracted from Getman \& Feigelson (2021). The filled polygons denote the solar flares (Feldman et al. 1995) and solar microflares (Shimizu 1995). The dashed lines indicate the 
    relation of $\mathrm{EM}\propto B^{-5}T^{17/2}$ for a constant magnetic field, and the dotted lines the relation of $\mathrm{EM}\propto L^{5/3}T^{8/3}$ at certain loop lengths (Shibata \& Yokoyama 1999).
    \label{fig:em-T}}
\end{figure*}

The event SVOM~J00365+0033 is marked on the EM$-T$ diagram (Shibata \& Yokoyama 2002) in Figure \ref{fig:em-T}, in which
the event falls within the upper-right area occupied by the RS CVn-type stars.

We estimate the loop length ($L$), electron density ($n_{\mathrm{e}}$) and magnetic field ($B$) by following the scaling laws given 
in Shibata \& Yokoyama (2002):

\begin{equation}
  L_9=f_{0.1}^{-3/5}\mathrm{EM}_{47}^{3/5}T_7^{-8/5}n_{09}^{-2/5}
\end{equation}
\begin{equation}
  n_9=10^{1.5}f_{0.1}^{2/5}\mathrm{EM}_{47}^{-2/5}T_7^{12/5}n_{09}^{3/5}
\end{equation}
\begin{equation}
  B_{50}=f_{0.1}^{1/5}\mathrm{EM}_{47}^{-1/5}T_7^{17/10}n_{09}^{3/10}
\end{equation}
where $L_9=L/10^9\ \mathrm{cm}$, $n_9=n/10^9\ \mathrm{cm^{-3}}$, $B_{50}=B/50\ \mathrm{G}$,
$\mathrm{EM}_{47}=\mathrm{EM}/10^{47}\ \mathrm{cm^{-3}}$, $T_7=T/10^7\ \mathrm{K}$ and $n_{09}=n_0/10^9\ \mathrm{cm^{-3}}$. $f_{0.1}=f/0.1$ is the filling factor of the hot plasma in the flare loop.
By adopting the typical value of $f_{0.1}=1$ and $n_{09}=1$, 
we find $L=1.7\times10^{12}\ \mathrm{cm}$, 
$n_{\mathrm{e}}=5.2\times10^9\ \mathrm{cm^{-3}}$ and $B=66$\ G. 

The properties estimated above are actually comparable to those of the superflares recently detected by LEIA in giant HD~251108 
(Mao et al. 2025), and by NuSTAR in NuSTAR\ J230059+5857.4 (Hakamata et al. 2025).    

\subsection{Plasma Motion Driven by Magnetic Reconnection}

Assuming the detected flare occurs in the primary of HD~22468\footnote{The primary K-type giant in
HD~22468 is believed to be more active than the secondary since the primary shows significantly stronger 
\ion{Mg}{2} emission lines during previous flares (e.g., Linsky et al. 1989).}, the escaping velocity can be
estimated to be $\upsilon_{\mathrm{esp}}=630(M_\star/M_\odot)^{1/2}(R_\star/R_\odot)^{-1/2}\approx
390\ \mathrm{km\ s^{-1}}$. Because this value is far larger than the outflow velocity determined 
from the modeling of the H$\alpha$ line profile, we suggest that the upflow revealed in this study can be 
interpreted by either chromospheric evaporation or prominence eruption. Both scenarios are related with 
the energy released in a magnetic reconnection process.

\subsubsection{Chromospheric Evaporation}

In the chromospheric evaporation scenario, the chromospheric plasma can be heated rapidly to a very high
temperature by Coulomb collisions of the electrons accelerated by the 
energy released in the magnetic reconnection (e.g., Fisher
et al. 1985; Canfield et al. 1990; Gunn
et al. 1994; Innes et al. 1997; Berdyugina et al. 1999; Li 2019; Tei et al. 2018; Yan et al. 2021; 
Fletcher et al. 2011; Tan et al. 2020; 
Chen et al. 2020). This heating results in an over-pressure in the chromosphere, 
which then pushes the plasma upward (e.g., Fisher et al. 1985; Teriaca et al. 2003; Zhang et al. 2016; Brosius \& Daw 2015; Tian \& Chen 2018) or downward (e.g., Kamio et al. 2005; Libbrecht et al. 2019; Graham et al. 2020).
Both upward and downward motion of chromospheric plasma have be infrequently diagnosed and reported 
for cool main-sequence stars and RS CVn-type stars in
previous studies (e.g., Honda et al. 2018; Koller et al. 2021; Wang et al. 2024; Cao \& Gu 2024, 2025). 

The upward velocity is typically of tens of kilometers per second in the ``gentle evaporation'' with 
an electron beaming flux $\leq10^{10}\ \mathrm{erg\ s^{-1}\ cm^{-2}}$ 
(Milligan et al. 2006; Sadykov et al. 2015; Li 2019). In addition, a fast upward motion with 
a velocity of hundreds of kilometers per second can result in 
an electron beaming flux $\geq3\times10^{10}\ \mathrm{erg\ cm^{-2}\ s^{-1}}$ (e.g.,
Milligan et al. 2006b; Brosius \& Inglis 2017; Li et al. 2017) in the case of a ``explosive evaporation''.

In the evaporation scenario, the mass of the moving plasma can be estimated
from the H$\alpha$ emission through the traditional method 
$M_{\mathrm{gas}}\geq N_{\mathrm{tot}}Vm_{\mathrm{H}}$ (e.g., Houdebine et al. 1990), where $N_{\mathrm{tot}}$ is the
number density of hydrogen atoms, $m_{\mathrm{H}}$ the mass of the hydrogen atom, and $V$ the total 
volume that can be determined from the line luminosity $L_{ji}$.
With $L_{ji}=N_jA_{ji}h\nu_{ji}VP_{\mathrm{esc}}$, 
the mass of the moving plasma $M_{\mathrm{PL}}$ can be estimated as

\begin{equation}
 M_{\mathrm{PL}}\geq\frac{4\pi d^2f_{\mathrm{line}}m_{\mathrm{H}}}{A_{ji}h\nu_{ji}VP_{\mathrm{esc}}}\frac{N_{\mathrm{tot}}}{N_j}
\end{equation}
where $N_j$ is the number
density of hydrogen atoms at excited level $j$, $A_{ji}$ the Einstein
coefficient for a spontaneous decay from level $j$ to $i$, $P_{\mathrm{esc}}$
the escape probability, $d$ the distance and $f_{\mathrm{line}}$ the observed line flux.

After removing the quiescent H$\alpha$ line emission and adopting $P_{\mathrm{esc}}=0.5$, the mass of the upward moving plasma 
is estimated to be as large as $M_{\mathrm{PL}} = 3.9\times10^{20}$g by transforming the H$\alpha$ line flux 
to that of H$\gamma$ by assuming a Balmer decrement of three (Butler et al. 1988).
$A_{52}=2.53\times10^6\ \mathrm{s^{-1}}$ (Wiese \& Fuhr 2009) 
and $N_{\mathrm{tot}}/N_5= 2\times 10^9$ are adopted in the calculation according to 
the non-local thermal equilibrium modeling by 
Houdebine \& Doyle (1994a, b). 

\subsubsection{Prominence Eruption}

Prominence eruption that has been detected in the decay phase of 
the flares in other stars and the Sun (e.g., Kurokawa et al. 1987) 
is an alternative explanation of the observed blueshifted H$\alpha$ emission 
(e.g., Otsu et al. 2022; Inoue et al. 2024; Wang et al. 2024). 

By following the method adopted in the previous studies (e.g., Maehara et al. 2021; 
Inoue et al. 2023; Wang et al. 2024), we estimate the mass of the prominence by 
\begin{equation}
  M_{\mathrm{p}}\approx 2m_{\mathrm{H}}\bigg(\frac{n_{\mathrm{H}}}{n_{\mathrm{e}}}\bigg)n_{\mathrm{e}}^{-1}\frac{d^2f_{\mathrm{H\alpha}}}{F_{\mathrm{H\alpha}}}\times \mathrm{EM}
\end{equation}
or numerically 
\begin{equation}
  \begin{split}
  M_{\mathrm{p}}\approx 3.2\times10^{19}\bigg(\frac{n_{\mathrm{H}}}{n_{\mathrm{e}}}\bigg)
  \bigg(\frac{n_{\mathrm{e}}}{10^{10}\ \mathrm{cm^{-3}}}\bigg)^{-1}
  \bigg(\frac{d}{\mathrm{pc}}\bigg)^2\times\\
  \bigg(\frac{f_{\mathrm{H\alpha}}}{10^{-10}\ \mathrm{erg\ s^{-1}\ cm^{-2}}}\bigg)
  \bigg(\frac{F_{\mathrm{H\alpha}}}{10^4\ \mathrm{erg\ s^{-1}\ cm^{-2}\ sr^{-1}}}\bigg)^{-1}\times\\
  \bigg(\mathrm{\frac{EM}{10^{30}\ \mathrm{cm^{-5}}}}\bigg)
  \end{split}
\end{equation}
where $m_{\mathrm{H}}$ is the mass of the hydrogen atom, $n_{\mathrm{H}}$ and $n_{\mathrm{e}}$ are 
the densities of hydrogen atom and electron, respectively. 
$d$ is the distance, and $f_{\mathrm{H\alpha}}$ the measured H$\alpha$ line flux,
$\mathrm{EM}=n_{\mathrm{e}}^2L$ is the emission measurement.
$F_{\mathrm{H\alpha}}$ is the prominence H$\alpha$ emission in unit of 
$\mathrm{erg\ s^{-1}\ cm^{-2}\ sr^{-1}}$, and depends on the optical depth.
As reported in Inoue et al. (2023),
the optical depth of the prominence H$\alpha$ emission is assumed to be 
$0.1<\tau<100$. 
The NLTE solar prominence model (Heinzel et al. 1994) gives
$F_{\mathrm{H\alpha}}\sim10^{4}\ \mathrm{erg\ s^{-1}\ cm^{-2}\ sr^{-1}}$ and 
$\mathrm{EM}\sim10^{28}\ \mathrm{cm^{-5}}$ in the case of $\tau=0.1$. 
In the case of $\tau=100$, the corresponding values 
are $F_{\mathrm{H\alpha}}\sim10^{6}\ \mathrm{erg\ s^{-1}\ cm^{-2}\ sr^{-1}}$
and $\mathrm{EM}\sim10^{31}\ \mathrm{cm^{-5}}$.

The properties of the flare estimated in Section 6.2 yields  
$\mathrm{EM}=n_{\mathrm{e}}L=4.6\times10^{31}\ \mathrm{cm^{-5}}$, which favors 
the optical thick case, and results in 
$M_{\mathrm{p}}\approx1.5\times10^{21}(n_{\mathrm{H}}/n_{\mathrm{e}})$. 
Taking into account the $n_{\mathrm{H}}/n_{\mathrm{e}}$ ratio ranging from 2.13 to 5.88 
for a prominence (Notsu et al. 2024), the prominence mass is estimated to be
$3.2\times10^{21}\ \mathrm{g}<M_{\mathrm{p}}<8.8\times10^{21}\ \mathrm{g}$ 
for the transient SVOM~J00365+0033.

\subsection{The Nature of the WL Flare}

Taking into account their special spectroscopic features and physical properties, 
three types of scenario have been proposed to understand the transporting of energy 
released in magnetic reconnection down to the lower chromosphere and photosphere
in RS CVn-type systems. 

An analogy with solar-type flares suggests that the 
WL flares in RS CVn-type systems are resulted from the process in which the particles accelerated 
in the magnetic reconnection bombard the upper photosphere. 
This scenario implies a violation of the the Neupert effect in SVOM~J00365+0033.
The Neupert effect, being first revealed in solar flares by Neupert (1968), refers to a significant relationship between the thermal coronal emission and the time-integrated non-thermal emission, 
in which the heating due to the accelerated particles leads to a rapid expansion into 
the coronal loop, i.e., the so-called chromospheric evaporation, and a gradual cooling back to the quiescent state through thermal emission\footnote{
The effect is commonly expressed as 
$F_{\mathrm{SXR}}\propto\int F_{\mathrm{HXR}}(t)\mathrm{d}t$ or $\mathrm{d}F_{\mathrm{SXR}}/\mathrm{d}t\propto F_{\mathrm{HXR}}(t)$, where 
$F_{\mathrm{SXR}}$ and $F_{\mathrm{HXR}}$ are the 
soft and hard X-ray flaring fluxes, respectively.} 
(e.g., Neupert 1968; Hudson \& Ohki, 1972; Antonucci et al. 1984). 
As the evidence supporting the solar-stellar flare connection, 
the Neupert effect has been frequently reported in 
previous studies for the flares on RS CVn-type stars, main-sequence late-type stars and T Tauri stars 
(e.g., Hawley et al. 1995, 2003; Guedel et al. 1996; Gudel et al. 2002a, b, 2004; Osten et al. 2004; Mitra-Kraev et al. 2005; Audard et al. 2007; Wargelin et al. 2008; Fuhrmeister et al. 2011; Lalitha et al. 2013; Caballero-Gereia et al. 2015; Tristan et al. 2023), based on 
multi-wavelength observations.


In a stellar flare,
its WL emission is commonly used as a proxy of the non-thermal hard X-ray ($\gg\mathrm{10 keV}$) 
emission, both because of the difficulty in
the detection of stellar hard X-ray emission and because the emission in
the two bands is found to be closely correlated in the solar flares (e.g., Kleint et al. 2016). 
Compared to the soft X-ray emission ($<20$~keV), the non-thermal emission traced by  
the WL emission is delayed by $\sim50$ minutes in SVOM~J00365+0033.

Some notable exceptions to the 
Neupert effect have actually been identified not only in solar flares, but also in some stellar flares 
(e.g., Doyle et al. 1998; Ayres et al. 2001; Osten et al. 2005),
although the underlying physics causing the breakdown of the Neupert effect is still an open issue.
Possible explanations include low ambient electron density (or EM), low efficiency of heating due to 
the high trapping efficiency, a deeper penetration of the photosphere resulting from more 
energetic particles and a transition region explosion with lower temperature (e.g., Ayres et al. 2001). However, the low EM (or electron density) scenario is unlikely to be supported by our estimation 
(see Sections 5.1 and 6.2).

The penetration length of the accelerated electrons is, however, an essential issue in applying the 
bombardment scenario to the WL flares in RS CVn-systems, although one could alternatively consider the role of beamed protons (e.g., Grinin \& Sobolev 1989).
The stopping length is predicted to be 
$h = 1.5\times10^{17}E^2/n_{\mathrm{e}}\ \mathrm{cm}$ for an electron with an energy $E$ in unit of keV
(Brown 1971), where $n_{\mathrm{e}}$ is the electron density in unit of $\mathrm{cm^{-3}}$. 
With the loop length ($L$) and $n_{\mathrm{e}}$ estimated above, the electrons with $E>170$~keV are 
required to penetrate down to the photosphere.

Because of this difficulty, other two scenarios, thermal conduction and shock waves, have been proposed 
to interpret the origin of the WL flares in RS CVn-systems (e.g., Mullan 1976; Foing et al. 1994; Paudel et al. 2021). Although Foing et al. (1994) preferred the conduction scenario for the 
exceptional WL flares on HD~22468 detected in 1989, the X-ray to optical energy ratio is 
estimated to be $\sim0.1$ in SVOM~J00365+0033, which is inconsistent with the value of unit adopted 
in Mullan (1976).

\section{Conclusions}

The transient SVOM~J00365+0033 is identified to be a superflare 
based on the associated WL flare and enhanced H$\alpha$ line emission.
The light curve and spectral analysis in both hard X-ray and optical bands allow us to arrive at following conclusions:
\begin{enumerate}
 \item The hard X-ray spectra of SVOM~J00365+0033 can be well reproduced by the \tt apec\rm\ model 
 with a high peak temperature of $106^{+27}_{-22}$ MK. The bolometric energy released in the event is estimated to be as high as 
 $\sim7.2\times10^{37}-1.7\times10^{38}\ \mathrm{erg}$.
 \item At $\sim$1.7 hrs after the trigger, an upward plasma motion with respect to the photosphere is 
 revealed by the bulk H$\alpha$ line blueshift at a velocity of $-96\pm20\ \mathrm{km\ s^{-1}}$.
 This motion could be caused by either a chromospheric evaporation or a prominence eruption. 

\end{enumerate}

%
\acknowledgments

We thank the anonymous referee for the helpful comments improving our study significantly.
The Space-based multi-band astronomical
Variable Objects Monitor (SVOM) is a joint
Chinese-French mission led by the Chinese National
Space Administration (CNSA), the French Space Agency
(CNES), and the Chinese Academy of Sciences (CAS).
We gratefully acknowledge the unwavering support of
NSSC, IAMCAS, XIOPM, NAOC, IHEP, CNES, CEA, and CNRS.
This study is supported by the National Key
R\&D Program of China (grant Nos. 2024YFA1611702, 2024YFA1611700), by the Strategic Pioneer Program on Space Science, Chinese Academy of
Sciences, grant No. XDB0550401, and by the National Natural
Science Foundation of China (grant Nos. 12494571, 12494570,12494573).
J.W. is supported by
the National Natural Science Foundation of China (under grants 12173009).  
We acknowledge the support of the staff of the Xinglong 2.16m telescope and SVOM/GWAC. 
This work was partially supported by the Open Project Program of the Key Laboratory of Optical Astronomy, National Astronomical Observatories, Chinese Academy of Sciences. 

\vspace{5mm}
\facilities{beijing: 2.16m}
\software{IRAF (Tody 1986, 1992), MATPLOTLIB (Hunter 2007), Heasoft,  Xspec (Arnaud 1996)}
          
\vskip12pt
\clearpage




\end{document}